\documentclass{article}
\usepackage{arxiv}
\usepackage{graphicx} % Required for inserting images
\usepackage{amssymb, amsmath}
\usepackage{bm}
\usepackage{xcolor}
\usepackage{multirow}
\usepackage{listings}
\usepackage{url}

\newcommand{\Qubit}[1]{\left|#1\right\rangle}
\newcommand{\ket}[1]{\left\langle#1\right|}
\newcommand{\NPr}[1]{\mathrm{Pr}\left(#1\right)}
\newcommand{\CPr}[2]{\mathrm{Pr}\left(#1\mid #2\right)}

\title{Representation of Classical Data on Quantum Computers}
\author{%
  Thomas Lang$^1$ \And
  Anja Heim$^1$ \And
  Kilian Dremel$^1$ \And 
  Dimitri Prjamkov$^1$ \And 
  Martin Blaimer$^1$ \And
  Markus Firsching$^1$ \And 
  Anastasia Papadaki$^1$ \And
  Stefan Kasperl$^1$ \And
  Theobald OJ Fuchs$^1$
}
\date{%
  $^1$Fraunhofer Institute of Integrated Circuits IIS, Development Center X-ray Technology\\[2ex]
  \today%
}

\begin{document}
\maketitle

\section{Motivation}
Quantum computing is currently gaining significant attention, not only from the academic community but also from industry, due to its potential applications across several fields for addressing complex problems~\cite{IndustryQC}.
For any practical problem which may be tackled using quantum computing, it is imperative to represent the data used onto a quantum computing 
system. Depending on the application, many different types of data and data structures occur, including regular numbers, higher-dimensional 
data structures, e.g., n-dimensional images, up to graphs. This report aims to provide an overview of existing methods for representing these data types on gate-based quantum computers.

\section{Gate-Based Quantum Computers vs. Quantum Annealers}\label{sec:annealing}
In theory, both gate-based quantum computers and quantum annealers are equivalent~\cite{GateAnnealEquiv}, yet in practice, 
currently only gate-based systems provide access to universal quantum computing, whereas annealers are typically exclusive to solving 
optimization problems. In the latter case, the datasets are encoded indirectly via its formulation within the Hamiltonian energy 
landscape as required by the adiabatic theorem~\cite{AdiabaticTheorem}, most often represented by a quadratic unconstrained optimization 
problem (QUBO)~\cite{QUBO}. This model allows only a binary representation of its variables, thus, for non-binary problems a proper 
embedding needs to be found as well. Conceptually, the following representation methods may also be used in this context. However, it is crucial to acknowledge that the related weighting terms might also need to be adjusted to preserve the original intent of the optimization problem. 

\section{Data Representation Methods}
The following paragraphs focus on general embeddings and representations which are proven to be effective on gate-based quantum systems. 

\subsection{Text}
Regarding the encoding of character sequences/strings, two main possibilities lie in either plainly converting it to a binary sequence 
(using a Unicode encoding, for example) or by applying a text embedding used in current machine learning applications. 
Concerning the latter, one possibility is to encode the strings in a categorical way, e.g., by using a Bag-of-Words approach~\cite{BoW}. 
There, strings are tokenized, and each word is recognized as an atom with an associated count of occurrences. The embedding then creates a 
high-dimensional numerical vector indicating for each token the count of occurrences. Similarly, machine learning-based techniques for 
converting sequences of strings to numeric vectors may be applied, such as word2vec~\cite{Word2Vec} or BERT~\cite{BERT}. 
Within word2vec, token streams of a training corpus are fed to a neural network, for which sliding windows consider a limited neighborhood 
of tokens at a time, with the intent of training the network to predict the current word given the context words in the neighborhood. 
After the training procedure, the internal weights of the neural network are used as a latent embedding for groups of tokens of the same 
size as during training, and these vectors of (real) numbers may again be encoded via the numerical vector encoding procedure described below.
The same procedure may be applied to the BERT model. 
 
\subsection{Categorical Values}
Considering categorical attributes, we can identify a set of different options like representations of genders (e.g., “male”, “female”, ...), 
cities and regional districts (e.g. “Nuremberg”, “Passau”, ...), and many more. In order to keep these categories fixed, i.e., to preserve 
their actual names, the values may be further embedded by a multitude of well-known strategies~\cite{Categorical}, among which ordinal encoding and 
one-hot encoding~\cite{OneHot} are the most widely utilized. For ordinal encoding, each category value is assigned a unique integer, 
which can be embedded as sketched below. Contrary, one-hot encoding creates vectors of indicators if a category is present or not. 
Furthermore, embeddings inferred by means of machine learning are applicable as well. 

Regardless of the specific method, all the mentioned embeddings should create numeric vectors, which can be represented on the quantum 
device as described below. 

\subsection{Numbers}
As for numerical (scalar) values, a binary representation is necessary, i.e., the representation as a bit-string similar to classical information encoding. Specifically, any given integer $x\in\mathbb{N}$ may be represented by its binary representation using $m$ bits via
$x=x_{m-1}\cdots x_0 = \sum_{i=0}^{m-1} x_i 2^i$, where the sequence $x_i\in\{0,1\}, i=0,\ldots,m-1$, forms the bit-string representation.
As bit-strings can be easily encoded on quantum devices when considering the computational basis $\{\Qubit{0},\Qubit{1},\ldots\}$, an
integer can be encoded by the tensor product state $\Qubit{x} = \Qubit{x_{m-1}\cdots x_0}$.
Constructing such a quantum state is achieved by flipping the according qubits using NOT gates. This concept can be also generalized 
to represent fixed-precision rational numbers~\cite{FloatingPoint}. 

It is easy to see that the basis encoding is \textbf{not} particularly efficient in terms of storage, as it still requires as many qubits 
as the bit-string of the given integer occupies, which is often even larger than the number of available qubits in currently available 
quantum devices.  
A more flexible approach for real numbers is by using the \emph{angle encoding} scheme. This technique considers any real number as the phase 
angle of a 1-qubit state in superposition. Specifically, if one considers some real value $x_i\in [0,2\pi]$ (or any real number, normalized to that interval), it can be encoded in the phase of a 1-qubit state via the embedding 
$x_i \mapsto \cos x_i \Qubit{0} + \sin x_i \Qubit{1}$~\cite{AngleEncoding}.
In an implementation, this state may be prepared by any desired rotation gate or sequences of them, while in practice most often the RY 
gate is used in conjunction with the computational basis. Thus, any real (normalized) number may be encoded in a 1-qubit state, which implies
that a value retrieval must be implemented accordingly. The reason for this is that the values of the amplitudes of the actual quantum state are different from the rotation angle in general when applying the exponential function. 
As an example, encoding the value $x_i=\pi$ via a rotation around the Y axis will result in $\Qubit{\pi}=RY(\pi)\Qubit{0}=\Qubit{1}$.
Naturally, this may be applied to integer values also, at the expense of losing exact arithmetic in practice. 

A representation of complex numbers is also possible. The quantum state of a single qubit can be represented by the superposition of two orthogonal basis states 
\[ 
\Qubit{\Psi} = \alpha \Qubit{0} + \beta \Qubit{1} = \cos{(\theta/2)} \Qubit{0} + e^{i\phi}\sin{(\theta/2)}\Qubit{1}
\] 
and thus, a complex number \( z = x + iy = \vert z \vert e^{i \phi} \) can be embedded using the angle encoding scheme by a rotation of \( \theta/2 \) along the Y axis with \( \theta = \vert z \vert \in [0, 2\pi] \) and an additional rotation along the Z axis by the phase angle \( \phi \). The rotation along two axes requires the rotation along the Y axis to be limited to \( \pi \) in radians. For that reason, \( \theta /2 \) appears in the sine and cosine terms. A geometric description of complex numbers embedded in a quantum state can be illustrated using the Bloch sphere. It should be noted that although complex numbers can be represented and processed on a quantum computer, only magnitude values can be retrieved by measuring the qubit quantum state.

\subsection{Vectors and Matrices}
In practical applications, most often data is not given as single scalar values but rather as collections of scalars, and therein most often as vectors. Fortunately, the above encoding schemes are also valid in this case by simply grouping the individual encodings using the Kronecker product. In case of the aforementioned angle encoding, a vector of data points $\bm{x}=\left(x_1,\ldots,x_n\right)^T\in\mathbb{R}^n$
can be embedded by the tensorization $\Qubit{\bm{x}} = \bigotimes_{i=1}^n R(x_i)\Qubit{0}$, where each scalar part is angle-encoded
by some rotation operator $R$ in a single 1-qubit state, and their (tensor product) combination forms the overall n-qubit state.

While conceptually simple, this angle encoding also has the benefit of requiring only a constant number of 1-qubit operations, 
which can be prepared in parallel and thus are optimal in terms of executing as many operations as possible within the decoherence time 
of the quantum device. However, a major caveat is that the quantum representation requires the \emph{same} number of qubits as the 
dimensionality of the input. To achieve a better memory compression of the input data, amplitude encoding was proposed instead for 
vectors $\bm{x}\in\mathbb{R}^{2^n}$ of exponential size $2^n$.
There, the input vector is encoded in a way such that its components form the measurement probabilities, i.e., the squared absolute 
values of the amplitudes, associated to a multi-qubit state. Since the latter need to sum up to one, the input vector is first normalized. 
Based on the superposition principle, one can now express the input vector via a n-qubit quantum state,
effectively distributing a sum having an exponential number of terms onto only linear qubits, whose tensorization captures the original
dimensionality~\cite{AmplitudeVector}. This is no longer possible in general with a simple preparation scheme, instead multiple approaches exist to perform this state preparation, e.g., the Quantum RAM approach~\cite{QRAM}. We want to note that amplitude encoding works only for vectors whose sizes are powers of two. If this is not the case, a vector may be padded with zeros to achieve that size. 

The embeddings for matrices can be easily achieved by stacking the columns of a matrix on top of each other, c.f., the mathematical 
vectorization method, and then embedding the resulting vector.

\subsection{N-Dimensional Images}
A particular interest lies in the representation of n-dimensional images. This not only concerns images as they are often generated with
imaging methods such as photography or X-ray computed tomography, but also many other domains where abstract data itself can be encoded
in a matrix/image, e.g., a time series of a fixed set of features where each row or column of an image may be a discrete time step.

It should be noted that retrieving images from a quantum state most often requires repeated measurements. This is due to the fact that most representations consider all pixels at once by entangling the position qubits. Consequently, a measurement typically only retrieves a single pixel. In order to retrieve the entire image from the quantum state, the complete quantum circuit needs to be executed many times, including the image preparation and any subsequent processing. Depending on the specific representation, this implies a theoretical infinite number of executions (e.g., for FRQI and QPIE) or at least as many shots so that each pixel is sampled at least once.

\subsubsection{FRQI}
One of the first examples of how to encode images on a quantum device is the \emph{Flexible Representation of Quantum Images} (FRQI)~\cite{FRQI}.
This first variant encodes any image by qubits representing the pixel position and exactly one qubit to represent any color/pixel value,
i.e., an $2^n\times2^n$ image $I$ whose pixel values $\theta_i, i=1,\ldots,2^{2n}-1$ are scaled to $[0,\pi/2]$ 
may be represented via the quantum state

\[ \Qubit{I} = \frac{1}{2^n} \sum_{i=0}^{2^{2n}-1} \left(\cos\theta_i\Qubit{0}+\sin\theta_i\Qubit{1}\right) \otimes \Qubit{i}, \]

where $\{\Qubit{0},\Qubit{1},\ldots,\Qubit{2^{2n}-1}\}$ is the computational basis.
We want to clarify that the normalization factor is indeed correct, i.e., $2^{-n}$ is correct instead of $2^{-2n}$ as one would 
intuitively think. This can be checked by verifying that $\|\Qubit{I}\| = 1$.

We see, that the pixel values are angle-encoded (after their normalization) and thus this method allows to store grayscale images or
color images alike, depending on the \emph{a priori} mapping to the interval $[0,\pi/2]$.
Additionally, it was proven that the number of simple gates necessary for preparing a FRQI state is polynomial in the dimension of the underlying Hilbert space (or equivalently exponential in the number of qubits)~\cite{FRQI}.

Note that while this is a one-dimensional scheme which represents the linearized image, its multi-dimensional equivalent can also 
be expressed by mapping the one-dimensional computational indices to multi-dimensional ones. We refer the reader to the 
example section for a demonstration.

\subsubsection{NEQR}
The discussed FRQI representation possesses some disadvantages, among which are a long image preparation phase, a non-exact retrieval
of pixels/images, and a non-convenience to express image processing operations. To counteract (some) of its disadvantages, the
\emph{Novel Enhanced Quantum Representation of digital images} (NEQR) was proposed. NEQR basically represents any image using
a superposition of position qubits, followed by an encoding of the color value of the "current" pixel. Thus, any $d$-bit image of size $2^n\times 2^n$
is expressed via
\[ \Qubit{I} = \frac{1}{2^n} \sum_{y=0}^{2^n-1} \sum_{x=0}^{2^n-1} \Qubit{f(x,y)}\Qubit{y}\Qubit{x}, \]
where $\Qubit{f(x,y)}$ is the quantum representation of the pixel value at position $(x,y)$.
Overall, $d+2n$ qubits are required to store such an image~\cite{NEQR}, which generalizes to $d+qn$ qubits for $q$-dimensional $d$-bit images
of size $2^n$ per dimension~\cite{NEQRICT}.

Due to its construction, the color and position values are stored in orthogonal dimensions and thus can be \emph{accurately} determined
using a projective measurement unlike the FRQI method.

Literature indicates that image preparation using NEQR can achieve an approximately quadratic speedup over the FRQI preparation procedure.
Additionally, a measurement of the quantum state retrieves an individual pixel, which encodes its position within the image and its value accurately (as opposed to probabilistic as in FRQI), allowing accurate image retrieval~\cite{NEQR}. Since each measurement retrieves a single pixel (accurately represented) with a certain probability, the expected values of the necessary number of measurements follows the order of $\mathcal{O}(2n 2^{2n})$ for images of size $2^n \times 2^n$, which forms a \emph{lower} bound on the number of measurements. Practical applications on simulators and real devices showed that this limit is approximately correct for ideal simulators. On the other hand, noise affection in real systems as well as decoherence issues severely impact the number of required measurements, which is in particular true for higher-dimensional images~\cite{NEQRICT}.

Color images may be represented in exactly the same way by extending the color qubits to the appropriate number.

\medskip
\textbf{GNEQR}\\
A generalized model of the NEQR representation encompasses the storage of rectangular images of size $2^m\times 2^n$. Given some function
$f(x,y)$ yielding the $d$-bit color value of the pixel at position $(x,y)$, the GNEQR representation~\cite{GNEQR} of this image is given by
\[ \frac{1}{\sqrt{2^{m+n}}} \sum_{y=0}^{2^n-1} \sum_{x=0}^{2^{m}-1} \Qubit{f(x,y)}\Qubit{y}\Qubit{x}. \]

\medskip
\textbf{QPIE}\\
A further extension of (G)NEQR is the \emph{Quantum Probability Image Encoding (QPIE)}~\cite{QPIE}. Given a 2D image of size $M \times  L$, a vector of size $ML$ can be formed. After normalizing the pixel values $f_k$ for each pixel $k$ to \( c_k = f_k/\|f\|_2 \), the image data can be mapped onto a quantum state of \(n =\log_2{(ML)} \) qubits:
\[
\Qubit{f}=\sum_{k=0}^{2^n-1} {c_k \Qubit{k}}.
\]
The computational basis $\Qubit{k}$ encodes the position $(x,y)$ of each voxel. Compared to (G)NEQR, there are no pixel value qubits required. Therefore, a $2^n \times 2^n$ image can be efficiently encoded by $2n$ qubits. The benefits of encoding an image of size $2^n \times 2^n$ is the possibility to explicitly express the row and column indices for QPIE~\cite{QPIEapplication}:
\[
\Qubit{f}=\sum_{y=0}^{2^L-1} \sum_{x=0}^{2^M-1}{c_{x, y} \Qubit{x} \Qubit{y}}.
\]
This expression allows the efficient application of quantum image transform algorithms. For example, a 2D quantum Fourier Transform (QFT) can be realized, by first applying a QFT to the row index qubits and then applying a QFT to the column index qubits. 
To extract useful information from QPIE embedded images, the quantum state of $\Qubit{f}$ will be measured several times. Each measurement will return a combination of qubits corresponding to the binary image index $i$ with probability $c_i$, hence the name quantum probability image encoding. 

The QPIE is useful in applications that require quantum image processing operations such as the 2D Fourier transform, the Hadamard transform, or the Haar wavelet transform. These operations can be efficiently integrated as subroutines and may provide exponential speedup over their classical counterparts. Therefore, QPIE is interesting for applications such as radio interferometry in astronomy~\cite{QPIEapplication} or magnetic resonance imaging (MRI), in which the inverse Fourier-transform is applied to obtain the final images.

\subsubsection{BRQI}
While popular image representations like FRQI and NEQR represent images by first arranging coordinates in the qubits followed by the color
information, the \emph{Bitplane Representation for Quantum Images} (BRQI) rearranges this by separating the individual bit-wise images.
Consider a $d$-bit image of size $2^k \times 2^n$. Therefore, each pixel value $f(x,y)$ at position $(x,y)$ is a $d$-bit value which can 
be represented by its binary representation $f(x,y) = v = v_{d-1}\ \ldots\ v_0$ with $v_i\in\{0,1\}, i=0,\ldots,d-1$.
The bitplane separation basically stores $d$ images after each other, where each image occupies only a single qubit for the current bit~\cite{BRQI}.
In mathematical terms, the $b$-th bitplane of such an image, i.e., a $2^k \times 2^n$ image where each color value is described by a 
single (classical) bit, is expressed in (generalized) NEQR format by
\[ \Qubit{I_b} = \frac{1}{\sqrt{2^{(n+k)}}} \sum_{x=0}^{2^k-1} \sum_{y=0}^{2^n-1} \Qubit{f(x,y,b)} \Qubit{x} \Qubit{y}, \]
where $f(x,y,b)$ denotes the $b$-th bit of the color value at position $(x,y)$. Opposed to GNEQR, BRQI rearranges the qubits such that the
outermost dimension corresponds to an enumeration of the bitplanes, yielding
\[ 
  \Qubit{I} 
    = \sum_{b=0}^{d-1} \Qubit{I_b} 
    = \frac{1}{\sqrt{2^{(n+k+\log_2(d))}}} \sum_{b=0}^{d-1} \sum_{x=0}^{2^k-1} \sum_{y=0}^{2^n-1} \Qubit{f(x,y,b)} \Qubit{x} \Qubit{y} \Qubit{b}.
\]
Thus, a BRQI image of the aforementioned dimension can be encoded using $k+n+\log_2(d)+1$ qubits, where the single qubit encodes the "binary color",
i.e., 1 or 0, for each bit.

Special care needs to be taken considering the number of bitplanes to make the $\log_2(d)$ be a well-defined term. In~\cite{BRQI}, this is limited
to $d=8$, having the advantage that $\log_2(8)=3$. This implies that all the color information in the image can be represented by $n+k+4$ qubits.
Compared to the $n+k+8$ qubits necessary for the GNEQR model, BRQI thus requires 4 qubits less, improving the storage efficiency by a factor of
$2^4=16$.

\subsubsection{QPIXL}
As the examples at the end of the paper will show, some of the aforementioned methods require a quite high number of qubits and
thus produce very high-dimensional state vectors. To overcome this limitation to some degree and to unify many of the above
methods, the \emph{quantum pixel representation} (QPIXL) was proposed. Its main goal is to provide a novel state preparation
procedure using multi-controlled RY gates, which afterwards may be decomposed in a sequence of RY and CX gates. In any way, first
a uniform superposition of all position qubits is established, followed by a unitary operation setting the color of the
"current" pixel. Depending on the implemented format, an angular encoding is used: For FRQI, this can be transferred directly, while
for NEQR, a constant angle is used whenever some bit in the color bitstring is 1, thus either requiring one or no multi-controlled 
RY gate~\cite{QPIXL}.

\subsubsection{Qutrit Representation}
A totally different approach resides in representing images using \emph{qutrits} instead of qubits. While a qubit is an instance of a 
two-dimensional Hilbert space, a qutrit is an element of a three-dimensional Hilbert space. It is thus represented as
$\alpha\Qubit{0}+\beta\Qubit{1}+\gamma\Qubit{2}$ under the usual condition that $|\alpha|^2+|\beta|^2+|\gamma|^2=1$.
Fundamentally, all operations on qutrits need to be reconsidered for such an approach\footnote{As must they for even more general qudit systems.},
including the Hadamard gate, so three basis states are in equal superposition. Since all such operations now consider three basis states at a time,
the base exponent changes, i.e., one considers $3^n\times 3^n$ images in this case~\cite{Qutrits}.

In a conceptually simple way, the FRQI representation can be translated to the qutrit case, which yields
\[ \Qubit{I} = \frac{1}{3^n} \sum_{i=0}^{3^{2n}-1} \left( \cos\theta_i\Qubit{0}+\sin\theta_i\Qubit{1}+0\Qubit{2} \right) \Qubit{i}. \]
Here, this is a direct translation for grayscale images which exploits only two out of three basis states in this formulation. Generalizations
to color images are available which store all RGB color parts in all three basis states by angle encoding~\cite{Qutrits}.
Furthermore, there also exists a qutrit version of the NEQR encoding for 8-bit images, given as
\[
 \Qubit{I} = \frac{1}{3^{n+1}} \sum_{b=0}^5 \sum_{i=0}^{3^{2n}-1} \Qubit{R_b^i G_b^i B_b^i} \Qubit{i},
\]
where $R_b^i$ is the $b$-th bit of the red color channel of the pixel at position $i$, likewise for $G_b^i$ and $B_b^i$ for the green and blue channels.
In case of grayscale values, one may simply omit the two other color channels.

We want to highlight the interesting upper limit for $b$ being five. This is due to the fact that the processing using qutrits enables to process
$3^2=9$ color bits simultaneously, while 8-bit colors only require six bits when expressed as ternary strings. Thus, the sum does not need to iterate
over all states there. Instead, one may encoding auxiliary information in it. However, the qutrit scheme seems to possess a better scaling capability
as compared to the regular FRQI qubit scheme~\cite{Qutrits}.

\subsubsection{Non-Standard Approaches}
We want to note that there exist further approaches to encode images on quantum devices, cf.~\cite{QImageSurvey}.
Two noteworthy examples are given using the \emph{Quantum Discrete Cosine Transform} (QDCT) and the \emph{Matrix Product State} (MPS) representation.

\textbf{QDCT}\\
The QDCT approach mimics the classical DCT for image encoding (as used in JPEG, for example) on a quantum device. The basic idea is to switch from the
image space to the frequency space, i.e., perform a mathematical basis transform. First, the image is stored via the NEQR scheme, then the 
transformed result typically gets truncated, effectively omitting high frequency parts consisting mainly of noise. 
Subsequent processing steps may serve the encryption of the image, additionally.
An according inverse transform is available to restore the original representation~\cite{QDCT}.

\textbf{MPS}\\
Taking the QDCT idea, a similar approach is found in MPS systems. MPS systems are typically used in quantum computing context for the increase of efficiency
in calculations by basically approximating a complex system (expressed as a operator) by a series of operators. The latter is computed by the Schmidt
decomposition (a generalization of the Singular Value Decomposition) and they may be truncated to discard "unimportant" information. The product of the
truncated operators (speaking with the idea of handling matrices in mind) yields a subspace approximation of the original input.
In terms of image storage, the idea is to express the image first in the Fourier basis (using the Discrete Fourier Transform), followed by a decomposition
into a MPS representation. Both FRQI and NEQR images may be input to this scheme~\cite{QImageMPS}.

\subsection{Time Series} 
An interesting part in many applications is the monitoring of processes, for which often sensor signals are recorded at discrete time steps
over the course of the object's usage period. Especially for the monitoring aspect, such \emph{time series} are of utmost importance.
Such a time series may be defined as $d\in\mathbb{N}$ signals measured at $n\in\mathbb{N}$ times, thus one may consider a 
$d$-dimensional vector of measurements $\bm{x}_t=\left(x_1,\ldots,x_d\right)^T$ at each time step $t=1,\ldots,n$.

Since we now basically deal with a set of vectors, the aforementioned vector encodings may be used directly. The temporal component will play a role in the
data processing later on (e.g., via auto-regressive models lifted to the quantum domain by using variational quantum circuits) rather than in the data encoding.

In case of missing data or unequal sampling distances, classical preprocessings may be applied. However, also quantum equivalents of
classical time series processing, e.g., exponential smoothing or forecasting, may be applied on quantum hardware as well~\cite{QTimeSeriesProc}.

For example, \cite{QTimeSeriesAmplitudeEncoding} processes (scalar-valued) time series data as follows: First, classical preprocessing is applied to resample
the time series to a length of $2^n$ entries. The encoding of the (normalized) vectors on $n$ qubits is performed by amplitude encoding
using a recursive scheme employing a series of rotation and controlled rotation gates. In detail, one starts with the first qubit $q_0$ and computes
the probability of the overall state having $q_0=\Qubit{1}$, yielding an angle for the encoding of the state 
$Pr(q_0=\Qubit{1})\Qubit{100} + (1-Pr(q_0=\Qubit{0}))\Qubit{000}$
conditioned on the first qubit only. The construction of that state can be achieved using a rotation gate by the angle $\theta=2\arcsin \sqrt{Pr(q_0=\Qubit{1})}$.
Then, each term of this state is further decomposed in the same manner, now fixing the first qubit and conditioning the next qubit in line.
That process is continued for all qubits. The result is a tree where each level spans the currently considered qubit and their respective probabilities,
each corresponding to an angle used in a (multi-qubit-)controlled rotation gate. 
A depth-first traversal of that tree yields the state preparation circuit.

Differently, \cite{QLSTM} encoded its input data by applying angular encoding to each scalar entry of each input vector. They report impressive prediction results when training a Quantum long short term memory network consisting of variational quantum circuits.

\subsection{Graphs}
Before describing different possibilities of representing graphs on gate-based quantum devices, it should be noted that due to their nature, many graph problems are usually expressed as an optimization problem and thus we would argue that typically quantum annealing is better suited to these types of problems, cf. Section~\ref{sec:annealing}.

On classical computers, several graph representations (adjacency matrix, adjacency list, etc) exist, and these are also translatable to the quantum case. Due to the high degree of sparsity in many graphs, the adjacency matrix representation is especially popular in the quantum domain, as this way any node and its local neighborhood can be assembled on the fly. 
An example is given in~\cite{GraphCNN}, where a graph convolutional neural network is implemented on gate-based quantum hardware. There, the local graph neighborhood - or more precisely their feature vectors - necessary for the convolution operation is encoded via angular encoding in the required qubits. The connection of those neighborhoods instead is employed in the implementation of the convolutional layer operation, which involves operations with controlled input qubits linked to the according nodes.

A more direct translation of an adjacency matrix in a quantum state is given in the formulation of \emph{graph states}. Within a graph state, each vertex is encoded as a single qubit and edges are translated by applying a unitary affecting only the relevant qubits, i.e., interpreting the graph as a multi-particle quantum system. Consider a graph $G=(V,E)$ as a tuple of vertices $V$ and edges $E = \{\{a,b\}\mid a,b\in V\}$. It should be noted that graphs come in many forms, be it directed vs. undirected, cyclical vs. acyclical, weighted vs. unweighted, and many more. Following~\cite{GraphStates}, we limit ourselves to undirected simple graphs, i.e., undirected graphs having neither loops (edges from a vertex to itself) nor multiple edges between the same vertices. Furthermore, we will consider weighted simple graphs, i.e., simple graphs where each undirected edge $\{a,b\}$ possesses a weight $\phi_{ab}$. For unweighted graphs, we have $\phi_{ab} = \mathrm{const}$.
Thus, the graph state representing $G$ is given by 
\begin{align*}
    \Qubit{G} = \prod_{\{a,b\}\in E} U_{ab}(\phi_{ab}) \Qubit{\Psi}^{\otimes|V|},
\end{align*}
where $\Psi$ is some initial state for each qubit/vertex and $U_{ab}(\phi_{ab})$ is a (non-local) unitary encoding the interaction between these qubits, representing the edge between the according vertices. The structure of simple graphs, or rather their adjacency matrix, namely symmetry and a non-fixed ordering, implies constraints on possible interactions, hence why~\cite{GraphStates} defines the according graph state of a simple graph using $\Qubit{\Psi} = \Qubit{+}$ and $U_{ab}(\phi_{ab}) = \exp{\left(-i\phi_{ab}\sigma_z^a\sigma_z^b\right)}$. Specifically for unweighted simple graphs, they propose to fix $\phi_{ab} = \pi$. These choices result in a maximally entangled state in which further $U_{ab}$ is designed to add the edge between vertices $a$ and $b$ or remove it via ${U_{ab}}^{\dagger}$. So, for an unweighted simple graph, the graph state simplifies to $\Qubit{G} = \prod_{\{a,b\}\in E} U_{ab}\Qubit{+}^{\otimes|V|}$.

For encoding directed graphs instead, \cite{EulerianMultiGraphs} proposed an encoding which basically converts the graph into a unitary matrix representing its connectivity, which simultaneously encodes the graph as a quantum circuit. The encoding procedure considers Eulerian multi-graphs (we refer the reader to~\cite{EulerianMultiGraphs} for an introduction), which are transformed via a two-step procedure into a unitary matrix. In detail, first a so-called line graph is constructed, representing an incidence matrix where subsequent edges in the graph are indicated by nonzero entries. For the next step, a family of unitary matrices is assumed a priori, each matrix corresponding to a certain possible input degree of any possible vertex. During the unitarization process, the graph vertices are visited and depending on its input-degree, a unitary matrix is selected which replaces according entries in the line graph adjacency matrix. We refer again to the paper for an example. We want to note that it further proposes an algorithm to transform any multi-graph into an Eulerian multi-graph, albeit this introduces new edges in general and thus may change the adjacency properties which needs to be adjusted using according projection operators.

\subsection{Product and Sum Types}
Completing our overview over different data encodings, we briefly discuss product types and sum types as they are known from several classical programming languages. Both are composite types, i.e., they combine several instances of potentially different types. Product types are used to denote collections of several variables which have to exist simultaneously, corresponding to the \texttt{struct} concept in many programming languages. Differently, sum types represent a kind of superposition of several variables, often accompanied by a tag indicating which variable is currently used, corresponding to the (tagged) \texttt{union} concept in classical programming languages.
In literature, there have been efforts in designing programming languages translating classical concepts to the quantum domain. \cite{QUnity} gives an overview over some approaches and itself proposes a new quantum programming language, including a formalism and denontational semantics for quantum sum and product types.

Consider the following schematic C code presenting a collection of two variables (\lstinline|var1,var2|) of respective types (\lstinline|T1,T2|), once as a product type / struct \lstinline|ST| whose members must exist at the same time, and once as a tagged union \lstinline|UN| where an according tag (outside of the union itself) identifies whether \lstinline|var1| or \lstinline|var2| is currently extant.
\begin{lstlisting}[language=C]
  struct ST {
    T1 var1;
    T2 var2;
  };

  union UN {
    T1 var1;
    T2 var2;
  };
\end{lstlisting}
In the quantum domain, the types \lstinline|T1| and \lstinline|T2| correspond to Hilbert spaces used to represent the variables, i.e., they are the Hilbert spaces in which the quantum states representing the variables reside.

As in product types both members need to exist simultaneously, the quantum equivalent of the above struct is given as a tuple $\Qubit{\mathrm{var1}} \otimes \Qubit{\mathrm{var2}}\in \mathrm{T1}\otimes\mathrm{T2}$. Consequently, the tensor product of all quantum representations of all struct members represents all members simultaneously, and the dimensionality and qubit requirements increase accordingly.

Contrary, sum types represent only one member at a time. \cite{QUnity} proposes a pattern matching syntax (and according semantics) to select that member which is deemed "active". Coarsely, they propose a syntactic matching producing
\begin{align*}
  \Qubit{\mathrm{UN}} \mapsto 
    \begin{cases}
      \Qubit{\mathrm{var1}}, &\mathrm{tag} = \mathrm{var1}, \\
      \Qubit{\mathrm{var2}}, &\mathrm{otherwise}.
    \end{cases}
\end{align*}
In the language semantics, they encode this by considering a superposition of all possible sum type members and associating to each member a special amplitude value, corresponding to the estimated probability that that state would be measured by an observer.

\section{Conclusion}
In this report, we summarized selected popular methods for representing different types of data on gate-based quantum devices. In general, these rely on the superposition principle to represent large quantities of data simultaneously. For example, the image encoding technique NEQR employs this method for generating a quantum state storing all pixels of an image at once, and subsequent algorithms may exploit this operating on all pixels, achieving a theoretical exponential save in "memory requirement". Using this idea, many types of data as they appear in industrial settings may be used in quantum applications, ranging from vector-based data, over images, time series, graphs, up to general product and sum types.
As a final remark, we want to note that any data representation must fit the algorithm processing that data. Thus, we expect that a lot more research for the efficient encoding of data of different types will be conducted alongside the development of new and better quantum processing algorithms.

\subsection*{Acknowledgements}
This work was supported by the German Federal Ministry of Economic Affairs and Climate Action on the basis of a decision by the German Bundestag through the project QuaST.

% ================================================================================================================================ %

\newpage
\appendix
\section{Examples}
The following paragraphs demonstrate exemplary the embeddings described above on toy examples. 

\subsection{Text}
As for a Bag-of-Words approach, consider the set of strings $\{$“I like this, but this is brilliant.”, “How about this?”$\}$. 
After a tokenization, the frequencies of all words (converted to lower case) are recorded, thus yielding several numeric vectors 
(here arranged as a table for clarity): 

\begin{tabular}{|c|c|c|c|c|c|c|c|c|}
  \hline
       & i & like & this & but & is & brilliant & how & about \\\hline
  str1 & 1 & 1 & 2 & 1 & 1 & 1 & 0 & 0 \\\hline
  str2 & 0 & 0 & 1 & 0 & 0 & 0 & 1 & 1 \\\hline
\end{tabular}

The resulting numeric vectors, e.g., $(1,1,2,1,1,1,0,0)^T$ for string 1, can be encoded on the quantum device by any vector embedding.

\subsection{Categorical Values}
Consider the category “Type of wine” with possible labels $\{$“red”, “rose”, “white”$\}$. An ordinal encoding assigns a unique integer 
to each label, i.e., it may produce a mapping $\{\text{"red"} \mapsto 0, \text{"rose"} \mapsto 1, \text{"white"} \mapsto 2\}$, 
and subsequent algorithms would encode the numeric values and do computations on them. One should be aware that such computations must 
still respect that these values have a special meaning and that, e.g., weighted averages may produce invalid or wrong labels. 
Therefore, one-hot encoding represents any assignment as indicator vectors, i.e., the label “red” is represented as vector $(1,0,0)^T$ 
while “white” is given by $(0,0,1)^T$. Thus, there is a way of differentiating the labels after the embedding.
Again, the generated vectors are numerical and can thus be embedded on the quantum device by any vector embedding.  

\subsection{Numbers}
An example of a basis encoding for an integer is rather trivial: Consider the integer 5 and its binary representation 101. 
The according quantum embedding corresponds to the state $\Qubit{101}$ and is thus given by a 3-qubit state 
$\Qubit{101}=\Qubit{1}\otimes\Qubit{0}\otimes\Qubit{1}=\left(0,0,0,0,0,1,0,0\right)^T$.

Regarding the angle encoding of some value, e.g., 42, we first need to rescale the value to the interval $[0,2\pi]$, which can be
done by converting the angle of $42^{\circ}$ to radians, yielding $\phi = 7\pi/30$. The subsequent angle embedding, considering a rotation
around the Y axis, yields the 1-qubit quantum state 
$$\Qubit{42^{\circ}}=\cos\phi\Qubit{0}+\sin\phi\Qubit{1}\approx 0.743\Qubit{0}+0.669\Qubit{1},$$
where we rounded the numeric coefficients to three digits after decimal.

For a complex number, e.g. 42 + i21, the angle encoding scheme can be applied as well. Converting the number to radians yields \( z = (42 + 21i)\pi/180 \approx 0.733 + 0.367i \) with \( \theta = \vert z \vert \approx 0.81 \) and phase angle \( \phi \approx 0.46 \) in radians.
This yields the following quantum state:
\[
\Qubit{\Psi} = \alpha \Qubit{0} + \beta \Qubit{1} = 
\cos{(\theta/2)} \Qubit{0} + e^{i\phi}\sin{(\theta/2)} \Qubit{1} 
\approx 0.917 \Qubit{0} + (0.365 + 0.178i) \Qubit{1}
\]
The quantum state measurement allows one to retrieve only the magnitude of z, because only the probabilities \( \vert \alpha \vert ^2 \) or \( \vert \beta \vert ^2 \) for finding the qubit to be in state \( \Qubit{0} \) or \( \Qubit{1} \), respectively, are accessible and hence, only the angle \( \theta = \vert z \vert \) can be measured.
 
\subsection{Vectors and Matrices}
As the angle embedding of a vector is particularly easy to determine given the information above, we leave this as an exercise to the reader. 

Instead, we give a brief example for an amplitude encoding of the vector $(1,3,0,1)^T$ of length $4=2^n$ for $n=2$.
Due to the derived $n=2$, it is immediately clear that the amplitude encoding constructs a 2-qubit state.
After normalization, we obtain $\frac{1}{\sqrt{11}}\left(1,3,0,1\right)^T$, 
which can be decomposed by its expansion in the computational basis into
\[
  \begin{bmatrix} 1 \\ 3 \\ 0 \\ 1 \end{bmatrix}
   = 1\begin{bmatrix} 1 \\ 0 \\ 0 \\ 0 \end{bmatrix}
   + 3\begin{bmatrix} 0 \\ 1 \\ 0 \\ 0 \end{bmatrix}
   + 0\begin{bmatrix} 0 \\ 0 \\ 1 \\ 0 \end{bmatrix}
   + 1\begin{bmatrix} 0 \\ 0 \\ 0 \\ 1 \end{bmatrix}.
\]
The final amplitude encoding results in the 2-qubit state $\frac{1}{\sqrt{11}} \left(1\Qubit{00}+3\Qubit{01}+1\Qubit{11}\right)^T$.

\subsection{N-Dimensional Images}\label{sec:ndimageex}
\begin{figure}
    \centering
    \includegraphics[width=.25\textwidth]{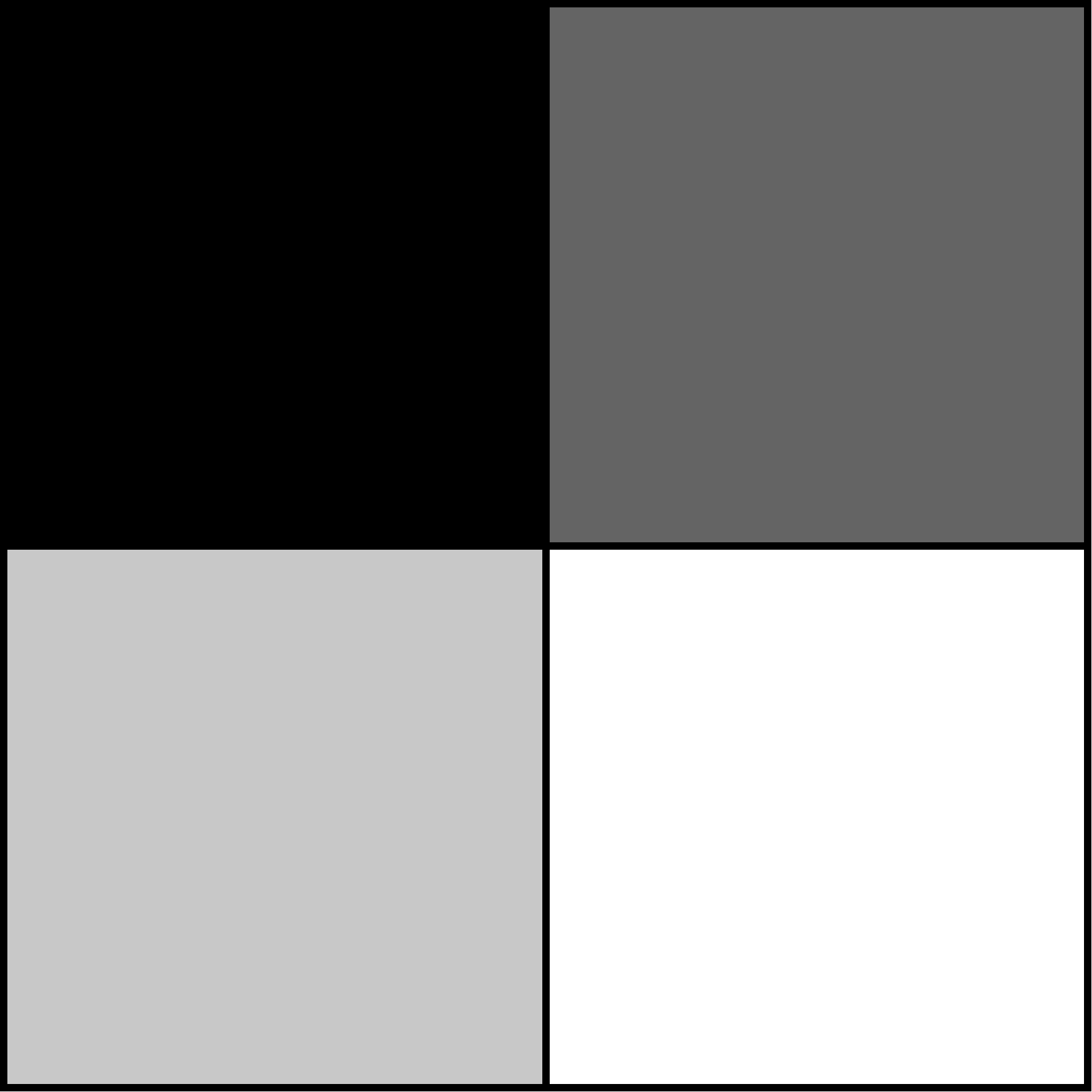}
    \caption{Illustrative 2-by-2 image to be encoded on a quantum device.}
    \label{fig:testimage}
\end{figure}

In these examples, we consider two-dimensional images for illustration purposes. However, please note that this may also be extended to
higher dimensions as well.
For all examples, consider a small 2 by 2 image (thus, $n=1$ in the following) given in Figure~\ref{fig:testimage}, 
whose grayscale values are 0 (black), 85 (dark gray), 170 (light gray) and 255 (white).

\subsubsection{FRQI}
First, let us consider an FRQI encoding. For that, we first need to rescale the grayscale values to the interval $[0,\pi/2]$, which is
reasonably easy since the maximum value is known. Thus, we map these values to angles by $x \mapsto \frac{x\ \pi}{255\cdot 2}$, yielding
according angles values of $0, \pi/6, \pi/3, \pi/2$.

To ease indexing, we also consider multi-dimensional computational basis vectors instead of one-dimensional ones, thus applying the following mapping:
\begin{align*}
    \Qubit{0} &\mapsto \Qubit{00} \\
    \Qubit{1} &\mapsto \Qubit{01} \\
    \Qubit{2} &\mapsto \Qubit{10} \\
    \Qubit{3} &\mapsto \Qubit{11} \\
\end{align*}

Combining both, the final FRQI state is given by
\begin{align*}
 \Qubit{I} 
     &= \frac{1}{2} 
              \left(
                  \left(\cos 0 \Qubit{0} + \sin 0 \Qubit{1}\right) \otimes \Qubit{00}
                + \left(\cos \pi/6 \Qubit{0} + \sin \pi/6 \Qubit{1}\right) \otimes \Qubit{01} \right.\\
                &\quad+ \left.\left(\cos \pi/3 \Qubit{0} + \sin \pi/3 \Qubit{1}\right) \otimes \Qubit{10}
                + \left(\cos \pi/2 \Qubit{0} + \sin \pi/2 \Qubit{1}\right) \otimes \Qubit{11}
              \right)\\
     &= \frac{1}{2} 
              \left(
                 \begin{bmatrix} 1\\0 \end{bmatrix} \otimes \begin{bmatrix} 1\\0\\0\\0 \end{bmatrix}
                +\begin{bmatrix} \cos\pi/6\\\sin\pi/6 \end{bmatrix} \otimes \begin{bmatrix} 0\\1\\0\\0 \end{bmatrix}
                +\begin{bmatrix} \cos\pi/3\\\sin\pi/3 \end{bmatrix} \otimes \begin{bmatrix} 0\\0\\1\\0 \end{bmatrix}
                +\begin{bmatrix} \cos\pi/2\\\sin\pi/2 \end{bmatrix} \otimes \begin{bmatrix} 0\\0\\0\\1 \end{bmatrix}
              \right) \\
     &= \frac{1}{2} 
       \begin{bmatrix}
         1 \\
         \cos\pi/6\\
         \cos\pi/3\\
         \cos\pi/2\\
         0\\
         \sin\pi/6\\
         \sin\pi/3\\
         \sin\pi/2
       \end{bmatrix}
\end{align*}

\subsubsection{NEQR}
Expressing the same image in the NEQR system, we obtain 
\[ 
  \Qubit{I} 
    = \frac{1}{2} 
        \left(
          \Qubit{00000000}\Qubit{00} + \Qubit{01010101}\Qubit{01} + \Qubit{10101010}\Qubit{10} + \Qubit{11111111}\Qubit{11}
        \right).
\]
We refrain from stating the actual vectors here, since this would require the printing of a 1024-dimensional vector.

\subsubsection{QPIE}
The first step in QPIE is the flattening of the image in Figure~\ref{fig:testimage} as a vector. After normalization with the L2-norm, we yield the corresponding pixel values: 
\[
\frac{1}{\sqrt{85^2 + 170^2 + 255^2}}[0, 85, 170, 255]^T = [0, 0.267..., 0.534..., 0.801...]^T
\]
This vector is then encoded as the superposition of n=2 qubits:
\[
\Qubit{I}= 0\Qubit{00}+0.267\Qubit{01} + 0.534\Qubit{10} + 0.801\Qubit{11}. 
\]

\subsubsection{BRQI}
Following the notion of the BRQI encoding, we first separate the image into its eight different bitplanes:

\begin{tabular}{|cc|cc|cc|cc|cc|cc|cc|cc|}
  \hline
    \multicolumn{2}{|c}{P0} & \multicolumn{2}{|c}{P1} & \multicolumn{2}{|c}{P2} & \multicolumn{2}{|c}{P3} 
  & \multicolumn{2}{|c}{P4} & \multicolumn{2}{|c}{P5} & \multicolumn{2}{|c}{P6} & \multicolumn{2}{|c|}{P7} \\\hline
  0 & 0 & 0 & 1 & 0 & 0 & 0 & 1 & 0 & 0 & 0 & 1 & 0 & 0 & 0 & 1 \\
  1 & 1 & 1 & 1 & 1 & 1 & 0 & 1 & 1 & 1 & 1 & 1 & 1 & 1 & 0 & 1 \\\hline
\end{tabular}

Now, we can represent each individual bitplane using the (G)NEQR representation, where the sum term combines all bit plane elements of the form $\Qubit{color}\Qubit{position}$, and the remaining three qubits are required to enumerate the bit planes.

\begin{alignat*}{3}
 \Qubit{P0}  &= \left(\Qubit{000}+\Qubit{001}+\Qubit{110}+\Qubit{111}\right)\otimes\Qubit{000} &&= [1,1,0,0,0,0,1,1]^T \otimes [1,0,0,0,0,0,0,0]^T \\
 \Qubit{P1}  &= \left(\Qubit{000}+\Qubit{101}+\Qubit{110}+\Qubit{111}\right)\otimes\Qubit{001} &&= [1,0,1,0,0,1,0,1]^T \otimes [0,1,0,0,0,0,0,0]^T \\
 \Qubit{P2}  &= \left(\Qubit{000}+\Qubit{001}+\Qubit{110}+\Qubit{111}\right)\otimes\Qubit{010} &&= [1,1,0,0,0,0,1,1]^T \otimes [0,0,1,0,0,0,0,0]^T \\
 \Qubit{P3}  &= \left(\Qubit{000}+\Qubit{101}+\Qubit{010}+\Qubit{111}\right)\otimes\Qubit{011} &&= [1,0,1,0,0,1,0,1]^T \otimes [0,0,0,1,0,0,0,0]^T \\
 \Qubit{P4}  &= \left(\Qubit{000}+\Qubit{001}+\Qubit{110}+\Qubit{111}\right)\otimes\Qubit{100} &&= [1,1,0,0,0,0,1,1]^T \otimes [0,0,0,0,1,0,0,0]^T \\
 \Qubit{P5}  &= \left(\Qubit{000}+\Qubit{101}+\Qubit{110}+\Qubit{111}\right)\otimes\Qubit{101} &&= [1,0,1,0,0,1,0,1]^T \otimes [0,0,0,0,0,1,0,0]^T \\
 \Qubit{P6}  &= \left(\Qubit{000}+\Qubit{001}+\Qubit{110}+\Qubit{111}\right)\otimes\Qubit{110} &&= [1,1,0,0,0,0,1,1]^T \otimes [0,0,0,0,0,0,1,0]^T \\
 \Qubit{P7}  &= \left(\Qubit{000}+\Qubit{101}+\Qubit{010}+\Qubit{111}\right)\otimes\Qubit{111} &&= [1,0,1,0,0,1,0,1]^T \otimes [0,0,0,0,0,0,0,1]^T 
\end{alignat*}

For the overall normalization constant, one obtains according to the BRQI specification $1/\sqrt{2^{(1+1+3)}} = 1/\sqrt{32}$.

Accumulating this yields the BRQI representation of the test image as
\begin{align*}
  \Qubit{I} = \frac{1}{\sqrt{32}} [11111111|10101010|01010101|00000000|00000000|01010101|10101010|11111111]^T
\end{align*}
where we left out the commas and used additional separators after each eighth entry for fitting this to the page width.

The advantage over NEQR can be seen immediately in the sense that BRQI requires a 64-dimensional vector (six qubits) for the representation of 
a $2\times 2$ image with 8-bit pixel values, whereas the same image in NEQR occupies a 1024-dimensional vector (ten qubits), thus taking up 16 times more
quantum resources than BRQI.

\subsubsection{Qutrit Representation}
Since qutrit images consider $3^n\times 3^n$ images, we encode the same image as above as the upper left corner of a $3\times 3$ image while all
other pixels are assumed to be zero. Following the NEQR-like encoding scheme, one actually iterates and sums over bitplanes for colors, where each "bit"
is actually a "trit". Thus, we first have to convert our decimal pixel values to ternary representation, yielding the following ternary image:

\begin{tabular}{|c|c|c|}
 \hline 
 000\ 000 & 010\ 011 & 000\ 000 \\\hline
 020\ 022 & 100\ 110 & 000\ 000 \\\hline
 000\ 000 & 000\ 000 & 000\ 000 \\\hline
\end{tabular}

Following the encoding scheme, we expand the encoding expression which sums over all bit planes and all pixels, i.e., for the $b$-th \emph{trit} of the color value at pixel $i\in\mathbb{N}$ denote the according state $\Qubit{C_b^i}\Qubit{i}$. As both systems live in respective Hilbert spaces, we actually represent both the pixel colors and the pixel positions in their ternary form. Furthermore, please note that the number of trits necessary for the color representation is six, as this is enough to represent all values in our range ($[0,255]$) in ternary form. As for the pixel position, we consider the linear index enumerating all nine pixels. Expanding the final sum yields
\begin{align*}
    \Qubit{I} 
        = \sum_{b=0}^d \sum_{i=0}^{3^{2n}-1} \Qubit{C_b^i}\Qubit{i}
        = \sum_{b=0}^5 \sum_{i=0}^8 \Qubit{C_b^i}\Qubit{i}
        = 3\Qubit{1}\Qubit{1} + 3\Qubit{2}\Qubit{3} + 3\Qubit{1}\Qubit{4}
\end{align*}
This results in a 81-dimensional real vector, the complete expansion is left as an exercise to the reader. Note that this way the image (and actually a larger image) can be represented by 4 qutrits while requiring 10 qubits for the same information.

\subsection{Time Series}
For a simple example, consider a time series consisting of four time points and two channels:
\[ 
 T=\left[ \begin{bmatrix}0.5\\0.8\end{bmatrix},\begin{bmatrix}0.3\\0.8\end{bmatrix},\begin{bmatrix}0.5\\0.9\end{bmatrix},\begin{bmatrix}0.45\\0.95\end{bmatrix} \right] 
\]
As we have two channels here, we concatenate them into a single vector with the first four entries corresponding to the first channel and the second four entries
corresponding to the second channel, respectively. Furthermore, we normalize the resulting overall vector, yielding
\[ T'= \left[ 0.26, 0.15, 0.26, 0.23, 0.41, 0.41, 0.46, 0.49 \right]^T \]
where all entries have been rounded to two decimal places for the sake of printing them here.
Thus, our overall goal is to encode this eight-dimensional vector as the quantum state
\[
\Qubit{T} = 0.26\Qubit{000}+ 0.15\Qubit{001}+ 0.26\Qubit{010}+ 0.23\Qubit{011}+ 0.41\Qubit{100}+ 0.41\Qubit{101}+ 0.46\Qubit{110}+ 0.49\Qubit{111}.
\]
requiring three qubits.

Following the encoding scheme in~\cite{QTimeSeriesAmplitudeEncoding}, we compute the amplitudes of the quantum states where only part of the qubits are fixed, 
e.g., we obtain $0.67\Qubit{11q_2}$ for a variable third qubit via $0.67=\|[0.49,0.46]\|_2$ from the states $0.49\Qubit{111}$ and $0.46\Qubit{110}$.
The following table shows the full computation:

\begin{tabular}{ccccccc}
\multirow{8}{*}{$\Qubit{q_0q_1q_2}$} 
 &\multirow{4}{*}{$\nearrow$}&\multirow{4}{*}{$0.89\Qubit{1q_1q_2}$}&\multirow{2}{*}{$\nearrow$}&\multirow{2}{*}{$0.67\Qubit{11q_2}$}&$\nearrow$&$0.49\Qubit{111}$ \\
 &                           &                                      &                           &                                    &$\searrow$&$0.46\Qubit{110}$ \\
 &                           &                                      &\multirow{2}{*}{$\searrow$}&\multirow{2}{*}{$0.58\Qubit{10q_2}$}&$\nearrow$&$0.41\Qubit{101}$ \\
 &                           &                                      &                           &                                    &$\searrow$&$0.41\Qubit{100}$ \\
 &\multirow{4}{*}{$\searrow$}&\multirow{4}{*}{$0.46\Qubit{0q_1q_2}$}&\multirow{2}{*}{$\nearrow$}&\multirow{2}{*}{$0.35\Qubit{01q_2}$}&$\nearrow$&$0.23\Qubit{011}$ \\
 &                           &                                      &                           &                                    &$\searrow$&$0.26\Qubit{010}$ \\
 &                           &                                      &\multirow{2}{*}{$\searrow$}&\multirow{2}{*}{$0.30\Qubit{00q_2}$}&$\nearrow$&$0.15\Qubit{001}$ \\
 &                           &                                      &                           &                                    &$\searrow$&$0.26\Qubit{000}$ 
\end{tabular}

Given this, the next goal is to compute the conditional probabilities of these "variable" states, e.g., $\CPr{q_1=\Qubit{1}}{q_0=\Qubit{1}}$. Since we know by
the given decomposition that the squared coefficients yields the probability of that state being measured, these conditional probabilities can be computed easily.
For example, we have that
\begin{align*}
 \NPr{q_0=\Qubit{1}} &= 0.89^2 = 0.79 \\
 \CPr{q_1=\Qubit{1}}{q_0=\Qubit{1}} &= \frac{\NPr{q_{01}=\Qubit{11}}}{\NPr{q_0=\Qubit{1}}} = \frac{0.67^2}{0.79} = 0.57  \\
 \CPr{q_2=\Qubit{1}}{q_{01}=\Qubit{11}} 
    &= \frac{\NPr{q=\Qubit{111}}}{\NPr{q_{01}=\Qubit{11}}} 
     = \frac{\NPr{q=\Qubit{111}}}{\CPr{q_1=\Qubit{1}}{q_0=\Qubit{1}}\NPr{q_0=\Qubit{1}}} 
     = \frac{0.49^2}{0.57 \cdot 0.79} = 0.53
\end{align*}
We proceed like this for all but the very last states in the above tree to obtain the following tree of conditional probabilities.

\begin{tabular}{ccccc}
 \multirow{4}{*}{$\NPr{q_0=\Qubit{1}} = 0.79$} & \multirow{2}{*}{$\nearrow$}
    &\multirow{2}{*}{$\CPr{q_1=\Qubit{1}}{q_0=\Qubit{1}} = 0.57$} & $\nearrow$ & $\CPr{q_2=\Qubit{1}}{q_{01}=\Qubit{11}} = 0.53$ \\
   &&                                                             & $\searrow$ & $\CPr{q_2=\Qubit{0}}{q_{01}=\Qubit{11}} = 0.47$ \\
   &\multirow{2}{*}{$\searrow$}
    &\multirow{2}{*}{$\CPr{q_1=\Qubit{0}}{q_0=\Qubit{1}} = 0.43$} & $\nearrow$ & $\CPr{q_2=\Qubit{1}}{q_{01}=\Qubit{10}} = 0.50$ \\
   &&                                                             & $\searrow$ & $\CPr{q_2=\Qubit{0}}{q_{01}=\Qubit{10}} = 0.50$ \\
 \multirow{4}{*}{$\NPr{q_0=\Qubit{0}} = 0.21$} & \multirow{2}{*}{$\nearrow$}
    &\multirow{2}{*}{$\CPr{q_1=\Qubit{1}}{q_0=\Qubit{0}} = 0.57$} & $\nearrow$ & $\CPr{q_2=\Qubit{1}}{q_{01}=\Qubit{01}} = 0.45$ \\
   &&                                                             & $\searrow$ & $\CPr{q_2=\Qubit{0}}{q_{01}=\Qubit{01}} = 0.55$ \\
   &\multirow{2}{*}{$\searrow$}
    &\multirow{2}{*}{$\CPr{q_1=\Qubit{0}}{q_0=\Qubit{0}} = 0.43$} & $\nearrow$ & $\CPr{q_2=\Qubit{1}}{q_{01}=\Qubit{00}} = 0.26$ \\
   &&                                                             & $\searrow$ & $\CPr{q_2=\Qubit{0}}{q_{01}=\Qubit{00}} = 0.74$ \\
\end{tabular}

For each of these (conditional) probabilities $p$, we can now compute an according rotation angle around the Y axis by $\theta=2\arcsin{\sqrt{p}}$.

Then, a depth-first traversal over the tree of states in conjunction with the according conditional probabilities yields a state preparation circuit
in which a controlled RY gate with the according angle is added. When the conditioning changes to the $\Qubit{0}$ qubit, according X gates are inserted.
The depth-first traversal visits the following encoding states in this order: 
$\Qubit{1\cdot\cdot}$,$\Qubit{11\cdot}$,$\Qubit{111}$,$\Qubit{101}$,$\Qubit{01\cdot}$,$\Qubit{011}$,$\Qubit{001}$.
There, the dots represent the "intermediate" states in the tree.

\begin{figure}
 \centering
 \includegraphics[width=\textwidth,trim=3cm 1.5cm 2cm 1cm]{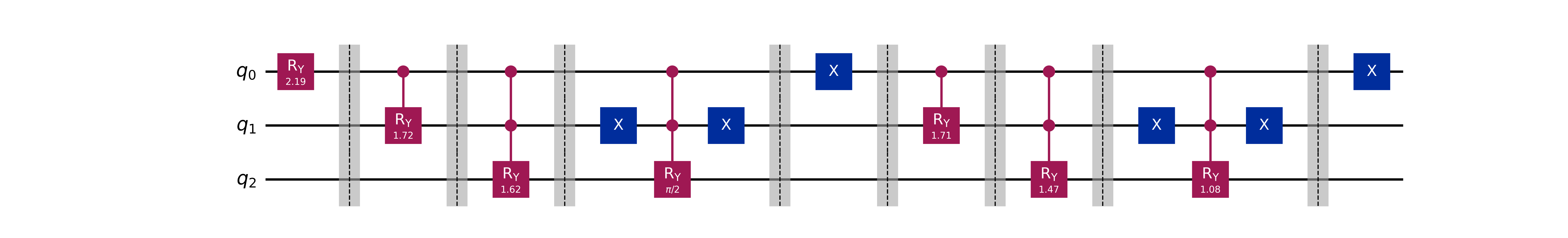}
 \caption{Quantum circuit for encoding a time series toy example.}
 \label{fig:timeseriesexample}
\end{figure}

In total, the circuit in Figure~\ref{fig:timeseriesexample} prepares a quantum state representing our example time series.

\subsection{Graphs}\label{sec:graphex}
In this example we consider the encoding of a simple graph consisting of three nodes ($V=\{1,2,3\}$) and two edges $E=\{\{1,2\},\{2,3\}\}$.
According to the graph state proposal, we first prepare the "empty" graph state 
\begin{align*}
  \Qubit{+}^{\otimes|V|} = \Qubit{+} \otimes \Qubit{+} \otimes \Qubit{+}
                         = \frac{1}{2\sqrt{2}} \begin{bmatrix}1\\1\\1\\1\\1\\1\\1\\1\end{bmatrix}.
\end{align*}
Next, we have to encode the edges as interactions between the qubits, and as described above we use the unweighted simple graph method of connecting them with a controlled Z gate on the according qubits. Following the idea found in~\cite{CZQubit}, we construct a "quantum if" statement, using $P_0 = \Qubit{0}\ket{0}$ and $P_1 = \Qubit{1}\ket{1}$, yielding
\begin{align*}
    U_{12} &= {CZ}_{12} = P_0\otimes I\otimes I + P_1\otimes Z\otimes I, \\
    U_{23} &= {CZ}_{23} = I\otimes P_0\otimes I + I\otimes P_1\otimes Z. 
\end{align*}
Therein, $U_{12}$ expresses that we apply the identity gate on qubit $3$ in any case. If the control qubit $1$ is in state $\Qubit{0}$, the first part of the sum is active and applies the identity transform on qubit $2$ as well, leaving the state intact. However, if the control qubit is in state $\Qubit{1}$ instead, the $Z$ gate is applied on qubit $2$ instead. 
A similar logic is applied in $U_{23}$, where the first qubit is left intact in any case and the phase of qubit $3$ is changed only if the control qubit $2$ is in state $\Qubit{1}$.

Expressing this in terms of matrices yields
\begin{align*}
  U_{12} &=   \begin{bmatrix}1&0\\0&0\end{bmatrix} \otimes \begin{bmatrix}1&0\\0& 1\end{bmatrix} \otimes \begin{bmatrix}1&0\\0&1\end{bmatrix}
            + \begin{bmatrix}0&0\\0&1\end{bmatrix} \otimes \begin{bmatrix}1&0\\0&-1\end{bmatrix} \otimes \begin{bmatrix}1&0\\0&1\end{bmatrix} \\
         &= \left[
              \begin{array}{cc|cc}
                 1 & 0                                   & \multicolumn{2}{c}{\multirow{2}{*}{0}} \\
                 0 & 1                                   & \multicolumn{2}{c}{}                   \\\hline
                 \multicolumn{2}{c|}{\multirow{2}{*}{0}} & \multicolumn{2}{c}{\multirow{2}{*}{0}} \\
                 \multicolumn{2}{c|}{}                   & \multicolumn{2}{c}{}                   \\
              \end{array}
            \right] \otimes \begin{bmatrix}1&0\\0&1\end{bmatrix}
            + \left[
              \begin{array}{cccc}
                 \multicolumn{2}{c|}{\multirow{2}{*}{0}} & \multicolumn{2}{c}{\multirow{2}{*}{0}} \\
                 \multicolumn{2}{c|}{}                   & \multicolumn{2}{c}{}                   \\\hline
                 \multicolumn{2}{c|}{\multirow{2}{*}{0}} & 1 &  0 \\
                 \multicolumn{2}{c|}{}                   & 0 & -1 \\
              \end{array}
            \right] \otimes \begin{bmatrix}1&0\\0&1\end{bmatrix} \\
         &= \begin{bmatrix}
               1 &   &   &   &   &   &    &     \\
                 & 1 &   &   &   &   &    &     \\
                 &   & 1 &   &   &   &    &     \\
                 &   &   & 1 &   &   &    &     \\
                 &   &   &   & 1 &   &    &     \\
                 &   &   &   &   & 1 &    &     \\
                 &   &   &   &   &   & -1 &     \\
                 &   &   &   &   &   &    & -1 
            \end{bmatrix}.
\end{align*}
Similarly,
\begin{align*}
  U_{23}  = \begin{bmatrix}
               1 &   &   &    &   &   &   &     \\
                 & 1 &   &    &   &   &   &     \\
                 &   & 1 &    &   &   &   &     \\
                 &   &   & -1 &   &   &   &     \\
                 &   &   &    & 1 &   &   &     \\
                 &   &   &    &   & 1 &   &     \\
                 &   &   &    &   &   & 1 &     \\
                 &   &   &    &   &   &   & -1 
            \end{bmatrix}.
\end{align*}

The encoding of the edges is finalized by applying them on the initial state, yielding the final graph state
\begin{align*}
  \Qubit{G} = U_{23}U_{12}\Qubit{+}^{\otimes|V|} 
    = \frac{1}{2\sqrt{2}}\begin{bmatrix}\phantom{+}1\\\phantom{+}1\\\phantom{+}1\\-1\\\phantom{+}1\\\phantom{+}1\\-1\\\phantom{+}1\end{bmatrix}.
\end{align*}

\subsection{Product Types}
As a simple example for product types, we consider the following example program:
\begin{lstlisting}[language=C++]
  struct ApplicationContext {
    Graph< Unweighted, Simple > graph;
    Image< 2, 2, uint8 > image;
  };

  Graph< Unweighted, Simple > graph;
  graph.vertices = {1, 2, 3};
  graph.edges    = { {1, 2}, {2, 3} };

  Image< 2, 2, uint8 > image;
  image[0, 0] =   0;
  image[0, 1] =  85;
  image[1, 0] = 170;
  image[1, 1] = 255;
  
  ApplicationContext ac;
  ac.graph = graph;
  ac.image = image;
\end{lstlisting}
As one can see, we consider an application requiring an unweighted simple graph and a $2\times 2$ 8-bit image. We consider the same examples as previously discussed in this work for the image (cf. Section~\ref{sec:ndimageex}) and the graph (cf. Section~\ref{sec:graphex}), respectively.

For the image, applying the FRQI encoding yields
\begin{align*}
  \Qubit{\mathrm{image}} =
    \frac{1}{2} 
       \begin{bmatrix}
         1 \\
         \cos\pi/6\\
         \cos\pi/3\\
         \cos\pi/2\\
         0\\
         \sin\pi/6\\
         \sin\pi/3\\
         \sin\pi/2
       \end{bmatrix},
\end{align*}
while the encoding of the graph yields the according graph state $\frac{1}{2\sqrt{2}}[1,1,1,-1,1,1,-1,1]^T$.

Combining these representations in a quantum product type creates their Kronecker product, and thus producing
\begin{align*}
  \Qubit{\mathrm{ac}} 
    &= \Qubit{\mathrm{graph}} \otimes \Qubit{\mathrm{image}} \\
    &= \frac{1}{4\sqrt{2}}\left[
        \begin{array}{r}
          \Qubit{I}\\\Qubit{I}\\\Qubit{I}\\-\Qubit{I}\\\Qubit{I}\\\Qubit{I}\\-\Qubit{I}\\\Qubit{I}
        \end{array}\right]\\
    &= \frac{1}{4\sqrt{2}}[\\
        &\qquad\qquad%
        \begin{array}{rrrrrrrr}
           1,  &  \cos(\pi/6), &  \cos(\pi/3), &  \cos(\pi/2), & 0, &  \sin(\pi/6), &  \sin(\pi/3), &  \sin(\pi/2), \\
           1,  &  \cos(\pi/6), &  \cos(\pi/3), &  \cos(\pi/2), & 0, &  \sin(\pi/6), &  \sin(\pi/3), &  \sin(\pi/2), \\
           1,  &  \cos(\pi/6), &  \cos(\pi/3), &  \cos(\pi/2), & 0, &  \sin(\pi/6), &  \sin(\pi/3), &  \sin(\pi/2), \\
           -1, & -\cos(\pi/6), & -\cos(\pi/3), & -\cos(\pi/2), & 0, & -\sin(\pi/6), & -\sin(\pi/3), & -\sin(\pi/2), \\
           1,  &  \cos(\pi/6), &  \cos(\pi/3), &  \cos(\pi/2), & 0, &  \sin(\pi/6), &  \sin(\pi/3), &  \sin(\pi/2), \\
           1,  &  \cos(\pi/6), &  \cos(\pi/3), &  \cos(\pi/2), & 0, &  \sin(\pi/6), &  \sin(\pi/3), &  \sin(\pi/2), \\
           -1, & -\cos(\pi/6), & -\cos(\pi/3), & -\cos(\pi/2), & 0, & -\sin(\pi/6), & -\sin(\pi/3), & -\sin(\pi/2), \\
           1,  &  \cos(\pi/6), &  \cos(\pi/3), &  \cos(\pi/2), & 0, &  \sin(\pi/6), &  \sin(\pi/3), &  \sin(\pi/2)\phantom{,}
         \end{array}\\
        &\qquad\qquad]^T
   % &= \frac{1}{4\sqrt{2}} 
   %      [
   %                      1,1,1,1,1,1,1,1,\\
   %        &\qquad\qquad \cos(\pi/6),\cos(\pi/6),\cos(\pi/6),\cos(\pi/6),\cos(\pi/6),\cos(\pi/6),\cos(\pi/6),\cos(\pi/6), \\
   %        &\qquad\qquad \cos(\pi/3),\cos(\pi/3),\cos(\pi/3),\cos(\pi/3),\cos(\pi/3),\cos(\pi/3),\cos(\pi/3),\cos(\pi/3), \\
   %        &\qquad\qquad \cos(\pi/2),\cos(\pi/2),\cos(\pi/2),\cos(\pi/2),\cos(\pi/2),\cos(\pi/2),\cos(\pi/2),\cos(\pi/2), \\
   %        &\qquad\qquad 0,0,0,0,0,0,0,0, \\
   %        &\qquad\qquad \sin(\pi/6),\sin(\pi/6),\sin(\pi/6),\sin(\pi/6),\sin(\pi/6),\sin(\pi/6),\sin(\pi/6),\sin(\pi/6) \\
   %        &\qquad\qquad \sin(\pi/3),\sin(\pi/3),\sin(\pi/3),\sin(\pi/3),\sin(\pi/3),\sin(\pi/3),\sin(\pi/3),\sin(\pi/3) \\
   %        &\qquad\qquad \sin(\pi/2),\sin(\pi/2),\sin(\pi/2),\sin(\pi/2),\sin(\pi/2),\sin(\pi/2),\sin(\pi/2),\sin(\pi/2) 
   %      ]^T
\end{align*}
as a vector in $\mathbb{R}^{64}$.

% =========================================================================================================================================== %

\end{document}